\documentstyle[aasms4]{article}
\input epsf
\begin{document}
\title{Supersonic Turbulence in the ISM: stellar extinction determinations
as probes of the structure and dynamics of dark clouds}

\author{Paolo Padoan}
\affil{Theoretical Astrophysics Center, Juliane Maries Vej 30, DK-2100 Copenhagen, Denmark}
\affil{Dept. of Astronomy, University of Padova, Vicolo dell' Osservatorio 5, I-35122, Padova, Italy}
\author{Bernard J. T. Jones}
\affil{Theoretical Astrophysics Center, Juliane Maries Vej 30, DK-2100 Copenhagen, Denmark}
\author{\AA ke P. Nordlund}
\affil{Astronomical Observatory and Theoretical Astrophysics Center, Juliane Maries Vej 30, DK-2100 Copenhagen, Denmark}
\authoremail{padoan@tac.dk}

\begin{abstract}

Lada et al. (1994) have described a method for studying the distribution of
dust in dark clouds using infrared imaging surveys. In particular they show
that the method provides some information about the structure of
the gas (dust) on scales smaller than their resolution.

In the present work we clarify the nature of the information provided
by their method.

We show that:

\begin{itemize}

\item the 3D density field of the gas is well described by a Log-Normal
distribution down to very small scales;

\item the power spectrum and the standard deviation of the 3D density
field can be constrained;

\item the origin of such a structure of the density field is likely to be
the supersonic turbulence in the gas.

\end{itemize}

In fact we find a qualitative and quantitative agreement between the predictions
based on recent numerical simulations of supersonic turbulence (Nordlund and
Padoan 1996; Padoan, Nordlund and Jones 1996) and the constraints
given by the infrared dust extinction measurements.

\end{abstract}


\section{Introduction}

In a recent paper Lada et al. (1994) have illustrated the method
of mapping the distribution of dust, and therefore gas, in dark clouds by
using stellar extinction measurements in the near-infrared. The method is
based on the use of multi-channel array cameras that allow the simultaneous
determinations of the colors of hundreds to thousands of stars through a
molecular cloud. The infrared color excess is proportional to the dust
column density, and the dust-to-gas ratio is known to be nearly constant in
interstellar clouds; therefore cloud maps can be obtained. The maps
obtained measuring the infrared excess are considerably more accurate than
the maps based purely on stellar counts.

The gas column density is given by

\begin{equation}
N(H+H_{2})=2\times 10^{21}A_{V} cm^{-2}
\label{1}
\end{equation}
where the visual extinction in magnitudes is
\begin{equation}
A_{V}=15.9E(H-K)
\label{2}
\end{equation}
and the color excess is
\begin{equation}
E(H-K)=(H-K)_{observed}-(H-K)_{intrinsic}
\label{3}
\end{equation}
and finally
\begin{equation}
<(H-K)_{intrinsic}>=13 \pm 0.01 mag.
\label{4}
\end{equation}
(see equations 1-4 in Lada et al. 1994).

The extinction data are used in two complementary ways, one
exploiting an ordered sampling (information on large scale
structure), the other a random sampling (information on scales
below the resolution of the map).

In the first method of analysis the data are spatially binned like in stellar count
and millimeter wave observations. At any position a few stars are
found so that an average extinction $A_{V}$ can be measured.
The result is an extinction map that compares well with the stellar
count map  and the CS map.

The second method is that of plotting the mean extinction, $A_{V}$,
and its standard deviation, $\sigma$, measured at any position. Lada
et al. (1994) found that the dispersion grows with the average
extinction, and realized that this behavior contains information
about the structure of the extinction (therefore of the gas mass
distribution) in the cloud, on scales smaller than the resolution
of the extinction map. They give examples of mass distributions
that would generate or not generate such a plot, but their interpretation
of the plot does not go very far.

In this work we focus on the second method of using the extinction data, that
is on the meaning of the $\sigma-A_{V}$ plot as a tracer of
structure on scales below the resolution of the map.

Fig.1, which is the equivalent of fig.7 in Lada et al. (1994), shows
the $\sigma-A_{V}$ plot obtained from the original data, kindly
provided to us by the authors. The measurements are taken for a
dark cloud complex near the young cluster IC 5146 in Cygnus. The
cloud has been mapped in $^{12}$CO and $^{13}$CO by Dobashi et
al. (1992), who named it `Cloud C'.

In sections 3 and 4 we show that the $\sigma-A_{V}$  plot is due to
the `intermittent' distribution of the dust (that is of the gas density
field in the cloud), and we show how to constrain such distribution
using randomly generated fields with given statistics and power
spectra. Before giving such details, though, we present  in the
next section the results of recent numerical simulations,
concerning the density field in isothermal  random supersonic flows.
It will be clear, in section 4 and in the following discussion, that
random supersonic flows are in fact excellent candidates to
interpret the extinction data and to explain the origin of the
distribution of dust in dark clouds.

\section{Supersonic Turbulence}

Nordlund and Padoan (1996) and Padoan, Nordlund and Jones (1996)
have recently discussed the importance of supersonic flows in shaping the
density distribution in the cold interstellar medium (ISM).

They have run numerical simulations of isothermal flows randomly forced
to high Mach numbers. Their experiments are meant to represent a
fraction of a giant molecular cloud: $\approx 10 pc$ in size and $10^3 M_{\odot}$
in mass. The simulated random supersonic motions are in fact observed in
molecular clouds.

It is found that most of the mass concentrates in a small fraction of the total
volume of the simulation, with a very intermittent distribution. The probability
density function (pdf) of the density field is well approximated by a
Log-Normal distribution:

\begin{equation}
P(lnx)dlnx=\frac{1}{(2\pi\sigma_{lnx}^{2})^{1/2}}exp\left[-\frac{1}{2}\left(\frac{lnx-\overline{lnx}}{\sigma_{lnx}}\right)^{2} \right]
\label{5}
\end{equation}
where $x$ is the relative number density:
\begin{equation}
x=n/ \overline{n}
\label{6}
\end{equation}
and the standard deviation $\sigma_{lnx}$ and the mean $\overline{lnx}$ are functions
of the rms Mach number of the flow, $\cal{M}$:

\begin{equation}
\overline{lnx}=-\frac{\sigma_{lnx}^{2}}{2}
\label{7}
\end{equation}
and
\begin{equation}
\sigma_{lnx}^{2}=ln(1+\frac{{\cal{M}}^{2}-1}{\beta})
\label{8}
\end{equation}
or for the linear density:

\begin{equation}
\sigma_{x}=\beta({\cal{M}}^{2}-1)^{0.5}
\label{9}
\end{equation}
where $\beta\approx0.5$.
Therefore the standard deviation grows linearly with the rms Mach number
of the flow.

It is also found that the power spectrum, $P(k)$, of the density distribution
is consistent with a power law:

\begin{equation}
P(k)\sim k^{-2.6}
\label{9bis}
\end{equation}
where $k$ is the wavenumber.

We will show in the following sections that the extinction data are consistent
with these theoretical predictions.

\section{Numerical Generation of Extinction Determinations}

In order to interpret the extinction data we have generated random 3D
density distributions with given statistics and power spectra, projected
them in 2D, and sampled them randomly as it happen when stars are found
through the cloud. The stars are assumed to be uniformly distributed
in space. Then a grid has been created on the distribution and the mean
extinction, $A_{V}$, and its dispersion, $\sigma$, have been measured in
every bin using the position selected by the few `stars' found in the bin.

In fig.2 we show the case of a Gaussian distribution, to be compared with the
case of a Log-Normal distribution, shown in fig.3. It is only in the case of the
Log-Normal distribution that the plot $\sigma-A_{V}$ is similar to the observational
one (fig.1). Clearly some sort of intermittent tail is needed in order to produce
the growth in the dispersion with the growth in mean extinction.

Intermittency is a natural explanation of the plot and it is also the main feature
of the density distribution in supersonic turbulence. Note that having only
high density clumps (e.g. steep power spectrum) is not enough to generate
the plot, as already shown by Lada et al. 1994.

We have studied the sensitivity of the plot to different power spectra and
standard deviations in the 3D density distribution. This is an important point because
the power spectrum cannot be measured accurately at the moment in the
numerical simulations of supersonic turbulence, since it requires a huge dynamical
range, and therefore it is interesting to find constraints for it via observations
of the projected density field in dark clouds with supersonic random motions.
The standard deviation is instead measured in the numerical simulations as a
function of the rms Mach number of the flow, ${\cal{M}}$, and may
be directly compared with the observed one, if the rms velocity in the cloud is
measured, as is done by millimeter wave observations.

Fig.4 shows a $\sigma-A_{V}$ plot from a random distribution with standard
deviation larger than the in the case of fig.3: the slope of the plot
increases together with the standard deviation of the density field.
This behavior is easily understood. In fact the plot is related
to the structure of the density field on a scale below the resolution of the
extinction map: if there were no structure on such small scale, $\sigma$
would be close to zero. The larger the fluctuations on small scale, the
larger $\sigma$.

\section{Statistics and Power Spectrum of the ISM Density Field}

Fig.1 shows the observational $\sigma-A_{V}$ plot. A linear regression
analysis gives:

\begin{equation}
\sigma=const + (0.35 \pm 0.01)A_{V}
\label{10}
\end{equation}
where the value of the constant is irrelevant in the present work, because
the numerical version of the plot can be freely translated along $A_{V}$.
Note that even the values $\sigma=0.0$ due to the presence of only a single
star in the bin are used. The elimination of those values would give the linear
regression coefficient found by Lada et al. (1994).

We want to understand now how the linear regression is affected by the
errors in the color excess. Lada et al. (1994) estimated a maximum error
in the color excess of $\pm 0.15$ mag, that translates into an error of
$2.5$ mag in $A_{V}$.

In order to study the effect of the color excess errors we have randomly added
such errors to the original data, both with a normal and with a uniform distributions.
This can be done many times, until any correlation $\sigma-A_{V}$ is
completely lost. By applying the errors once, we find on the average a
coefficient $(0.33\pm 0.02)$. This is an encouraging result, because it
means that, even after the addition of the errors, the uncertainty in the
coefficient is still low. Of course the coefficient has decreased a bit, because
the correlation between $\sigma$ and $A_{V}$ is diminished every time
errors are added.

The $\sigma-A_{V}$ relation to be compared with the numerical ones is
therefore:

\begin{equation}
\sigma=const + (0.36 \pm 0.02)A_{V}
\label{11}
\end{equation}

\subsection{Statistics}

As we mentioned in the previous section the slope of the numerical plot
depends on both the power spectrum index, $\alpha$, and the standard
deviation of the 3-D density distribution, $\sigma_{x,3D}$. We can
therefore draw lines of constant $\sigma-A_{V}$ linear regression
coefficient, $C_{r}$, on the plane $\alpha-\sigma_{x,3D}$, as shown in
fig.5.

Since the value of $C_{r}$ is known observationally with very small
uncertainty (see (\ref{11})), its contours on the numerical plane
$\alpha-\sigma_{x,3D}$ may in principle be used to constrain the power
index of the 3-D density field when its standard deviation is known,
or vice-versa.

In fact fig.5 shows that the lines of constant $C_{r}$ are almost
lines of constant $\sigma_{x,3D}$, which means that the plane
$\alpha-\sigma_{x,3D}$ is can constrain the value of
$\sigma_{x,3D}$, but not that of $\alpha$. Given the observational value of
$C_{r}$ and the value of $\alpha$ determined below, one gets:

\begin{equation}
\sigma_{x,3D}=5.0\pm 0.5
\label{12}
\end{equation}

This is the value of the standard deviation of the 3-D density distribution
in the `Cloud C', as given by stellar extinction measurements.

\subsection{Power spectrum}

The $\alpha-\sigma_{x,3D}$ plane does not constrain directly the index
of the (power law) power spectrum of the 3-D density distribution. Nevertheless it is
indirectly useful because it gives the 3-D standard deviation, that can
be compared with the observed 2-D standard deviation, in order to constrain
the spectral index. In fact the projection into 2-D of the 3-D distribution
is such that the two standard deviations are related in a way that depends
on the value of the power spectrum index. This can be shown using the
numerically generated random distributions. In fig.6, lines of constant
spectral index are plotted in the plane $\sigma_{x,2D}-\sigma_{x,3D}$.
For a fixed 3-D standard deviation, the value of the projected
2-D standard deviation decreases towards steeper spectra.

The 2-D standard deviation, $\sigma_{x,2D}$, is measured in the extinction
map, on a regular grid that contains on average about 5 stars per
bin. Its standard deviation is:

\begin{equation}
\sigma_{x,2D}=0.7\pm 0.1
\label{13}
\end{equation}

Entering the plane $\sigma_{x,2D}-\sigma_{x,3D}$ with this value and with
the previously determined value of $\sigma_{x,3D}$, one gets:

\begin{equation}
\alpha=-2.6 \pm 0.5
\label{14}
\end{equation}

This is the value of the power index of the 3-D density distribution in the
`Cloud C'.

\section{Discussion}

We have seen that the origin of the $\sigma-A_{V}$ plot is the intermittency
in the 3-D density distribution of the dark cloud; i.e., the
occurrence of huge density fluctuations with a significant probability.
We were inspired towards this explanation of the plot by recent results
emerging from our numerical experiments of highly supersonic turbulence.
The numerical experiments showed that most of the mass concentrates in
a small fraction of the total volume, that very large (orders of magnitude)
density contrasts appear in the flow, that the distribution of mass density
is well described by a Log-Normal, and that the standard deviation of
the statistics grows linearly with the rms Mach number of the flow.

Nevertheless the extinction map itself shows that the distribution of the
projected density has an intermittent tail that resembles a Log-Normal,
and even the distribution based on the sampling of column density star by
star, which is a random sampling, is qualitatively the same.

The problem here is a typical one in astronomy: extracting whole 3-D fields
from their integrated 1-D (eg velocity fields at any point) or 2-D
(eg scalar fields in space) counterparts that are observed from our fixed
point of view. One way to do this is to build a model for the 3-D
field and simulate the observational procedure on that field. The result
of the 'numerical' observation is compared with the actual observation, and
one may conclude whether or not the model field is consistent with the observation.
This method is good only as far as the observations are difficult to
reproduce, that is as far as there is no more than one reasonable model
that fits the observational data.

We have shown how the generation of random fields with given statistics and
power spectra (assumed to be power laws) leads to the plot of contours
of constant value of the slope of the $\sigma-A_{V}$ relation, on the plane
$\alpha-\sigma_{x,3D}$, and how one may extract both standard deviation and
spectral index of the original 3-D field of the
dark cloud.

Clearly {\it the contours of $C_{r}$ on the plane $\alpha-\sigma_{x,3D}$, obtained
with numerically generated random distributions, is a powerful tool
to investigate the 3-D structure of a dark cloud, down to scales smaller than
the resolution of the extinction map, when stellar extinction determinations
through the cloud are available.}

If it is assumed that the power spectrum is a universal property of
turbulence in this regime (highly supersonic and super-Alfv\'{e}nic,
isothermal equation of state), that its shape is in fact a power law, and that
dark clouds are in fact in such a turbulent state (as this work indicates)
then the observational constraint on the power spectrum obtained in the present
work is a prediction that numerical simulations will be able to check more
firmly in a few
years, when larger numerical simulations (N=500$^3$-1000$^3$) will be
available. In the meanwhile, this prediction is very useful for order to
model the origin of the mass distribution of protostars, given the statistics
and the power spectrum of the density field, in a similar way as it is done in
cosmology for predicting the mass distribution of galaxies (Press \& Schecther 1974).
This has in fact been done with considerable success (Padoan, Nordlund, \&
Jones 1996).

It should be noted that the method of using dust extinction measurements
in order to constrain the 3-D density field of dark clouds has several
advantages, when compared with the traditional method of using maps of molecular
emission lines.

First of all it is a much smaller observational effort.

Secondly the translation of the flux at one given line of one given molecule
into column density through the whole cloud is far more complex than the
transformation of stellar color excess into column density.

Finally the random sampling of points in space (random locations
of single stars behind the cloud) allows for the extraction
of information from scales smaller than the resolution of
the extinction map, based on averaging at any position a few of the
randomly selected points. The information is extracted by using the
observational $\sigma-A_{V}$ plot, together with the numerical prediction
of the slope of that plot in the $\alpha-\sigma_{x,3D}$ plane.

The extra information one gets can be summarized in the following points:

\begin{itemize}
\item there must be structure on scales at least ten times smaller than the
resolution of the extinction map, that is down to ~0.02 pc.
\item the density distribution is consistent with a Log-Normal
\item standard deviation and power index are measured
\end{itemize}

The present work suggests that the dust extinction measurements are
consistent with a scenario where the origin of the complex density field
in dark clouds is supersonic turbulence, or more generally the presence
of supersonic motions in dark clouds, whatever their
origin might be. Such a scenario is appealing, for the simple reason
that supersonic motions have indeed been observed and measured in dark
clouds for the last twenty years!

The connection between the observations and supersonic
turbulence is not only qualitative in nature (Log-Normal shape of the
3-D density distribution), but also quantitative.

In fact, in our numerical experiments (Nordlund \& Padoan 1996) we have
determined the relation between the standard deviation of the 3-D
density field, $\sigma_{x,3D}$, and the rms Mach number of the flow,
${\cal {M}}$ cf. Eq.\ (\ref{9}). The same may
be done observationally, since Cloud C has been studied in some
details by Dobashi et al (1992) in $^{12}$CO and $^{13}$CO.
They obtain measurements of temperature and velocity dispersion
in the three main cores, C1, C2 and C3, that are included in the area
where stellar extinction is measured by Lada et al. (1994). Using those
values one gets ${\cal {M}}\approx 10$. Given the value
$\sigma_{x,3D}=5$ obtained in the present work, one sees that the relation
between standard deviation of the density field and rms Mach number of the
flow is consistent with the one predicted in our numerical experiments
of supersonic turbulence.

Moreover, the spectral index estimated from our simulations,
$\sim -2.6$, is consistent with that obtained
from the observational data (see fig.7).

The values of the standard deviation and of the spectral index are useful
for modeling the origin of the mass distribution of protostars, given the statistics
and the power spectrum of the density field, in a similar way as it is done in
cosmology for predicting the mass distribution of galaxies (Press \& Schecther 1974).
This has in fact been done with considerable success (Padoan, Nordlund, \&
Jones 1996).

\section{Conclusions}

In the present work we have re-interpreted the
observational results obtained by Lada et al. (1994), i.e. the fact that
the mean stellar extinction at any given position in space increases together
with the dispersion of the extinction, where the averages are taken among the
stars found at that position in space.

The authors were able to conclude that:

\begin{itemize}
\item structure must be present down to scales smaller that the extinction
map resolution;
\item generic models for the cloud structure (eg uniform or in clumps) do not
easily reproduce the $\sigma-A_{V}$ plot;
\item the $\sigma-A_{V}$ plot is a basic test for any model for the dynamics
and structure of the cold interstellar medium.
\end{itemize}

We have simulated the observations by generating random density distributions.
In this way we have been able to better define the information contained
in the $\sigma-A_{V}$ plot. We can therefore add that:

\begin{itemize}
\item the statistics of the 3-D density field in the dark cloud is certainly
very intermittent; in particular it is consistent with a Log-Normal
distribution;
\item the standard deviation of the statistics is $\sigma_{x,3D}=5.0\pm0.5$;
\item the index of the power spectrum (assumed to be a power law), of the 3-D
density field, is $\alpha=2.6\pm0.5$;
\item the relation between the rms Mach number of the flow and the standard
deviation of the 3-D density field is about $\sigma_{x,3D}\approx0.5{\cal{M}}$;
\end{itemize}

We therefore conclude that the scenario for star formation and for the
dynamics of dark clouds proposed by Padoan (1995), Nordlund  \& Padoan
1996), and Padoan, Nordlund, \& Jones (1996) is fully consistent with the dust
extinction measurements in `Cloud C' by Lada et al. (1994).

In that scenario the dynamics of dark clouds is characterized by supersonic
random motions, which are responsible for fragmenting the mass distribution.
In the cited works we showed that a MIller-Scalo stellar mass function is
a natural consequence of that scenario. Here we have shown that the
shape, the standard deviation, and the spectral index of the density distribution,
predicted with numerical simulations of supersonic turbulence and used in that
scenario, are consistent with the observations of stellar extinction.

\acknowledgements

We thank Charles J. Lada for pointing our attention on the observational
results and for kindly providing the original data.

This work has been supported by the Danish National Research Foundation
through its establishment of the Theoretical Astrophysics Center.

\newpage
\begin{figure}
\centering
\leavevmode
\epsfxsize=1.0
\columnwidth
\epsfbox{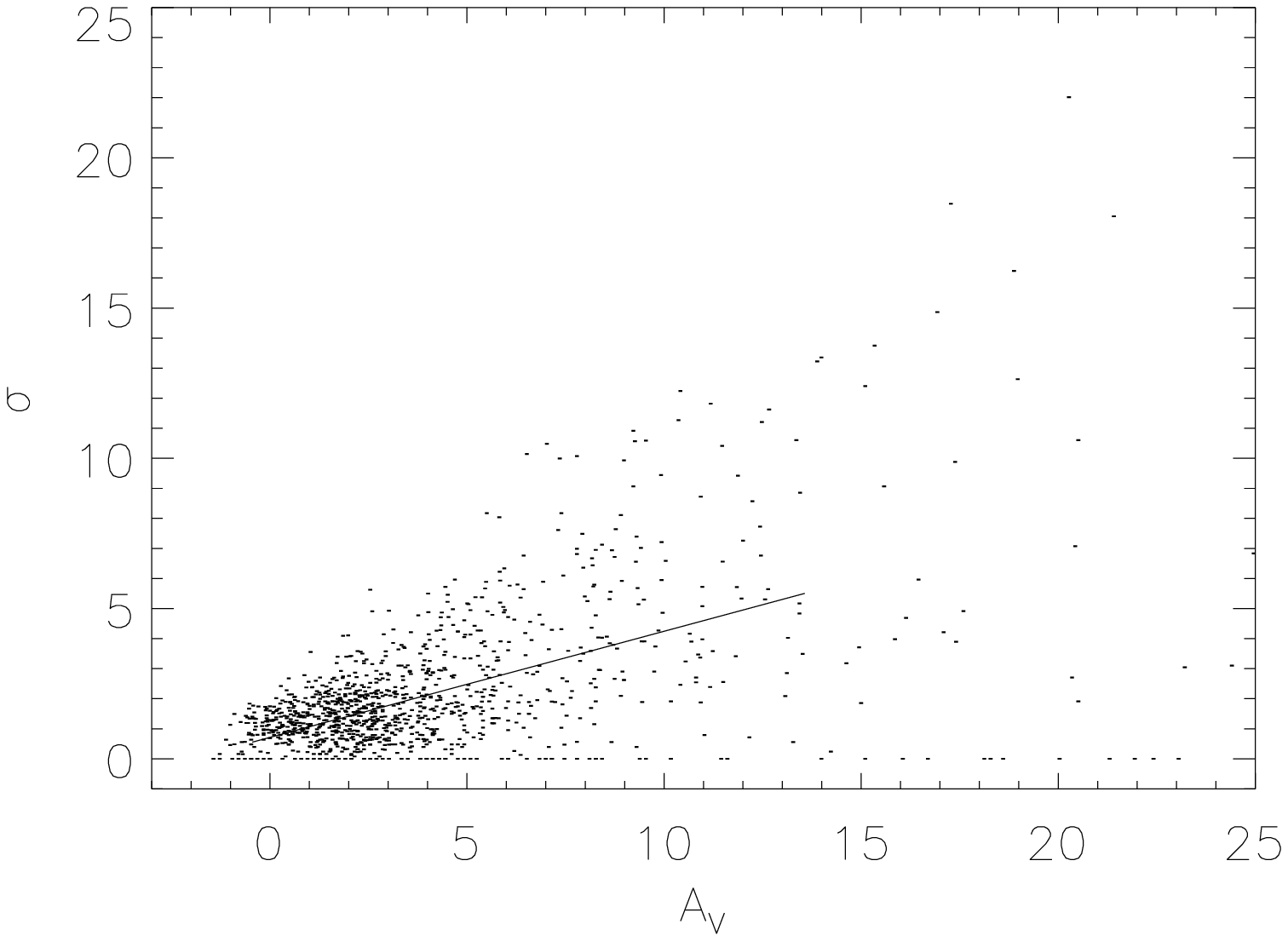}
\caption[]{The dispersion versus the mean extinction for every bin
in the regular grid superposed to the observed region. A bin
contains on average about 5 stars. The data are the original ones
from Lada et al. (1994)}
\end{figure}

\newpage
\begin{figure}
\centering
\leavevmode
\epsfxsize=1.0
\columnwidth
\epsfbox{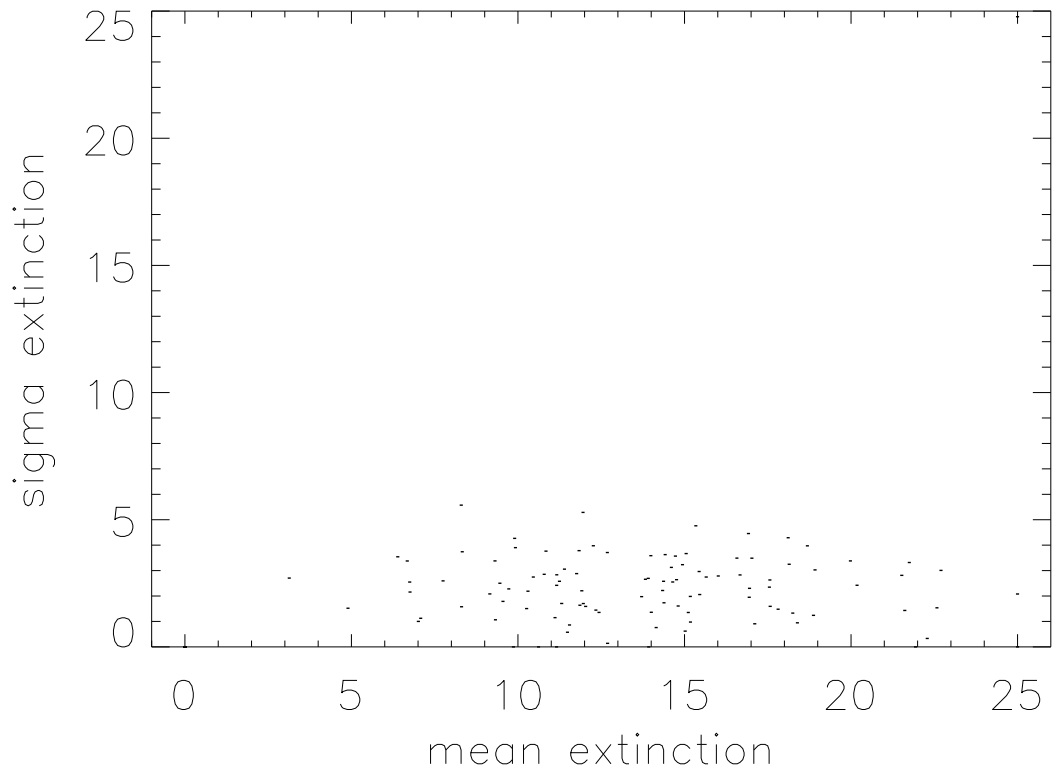}
\caption[]{The same plot as in fig.1, but obtained numerically starting from
a 3-D random distribution with a Gaussian statistic. The Gaussian
statistic is clearly unable to reproduce the observed growth of dispersion with
mean extinction.}
\end{figure}

\newpage
\begin{figure}
\centering
\leavevmode
\epsfxsize=1.0
\columnwidth
\epsfbox{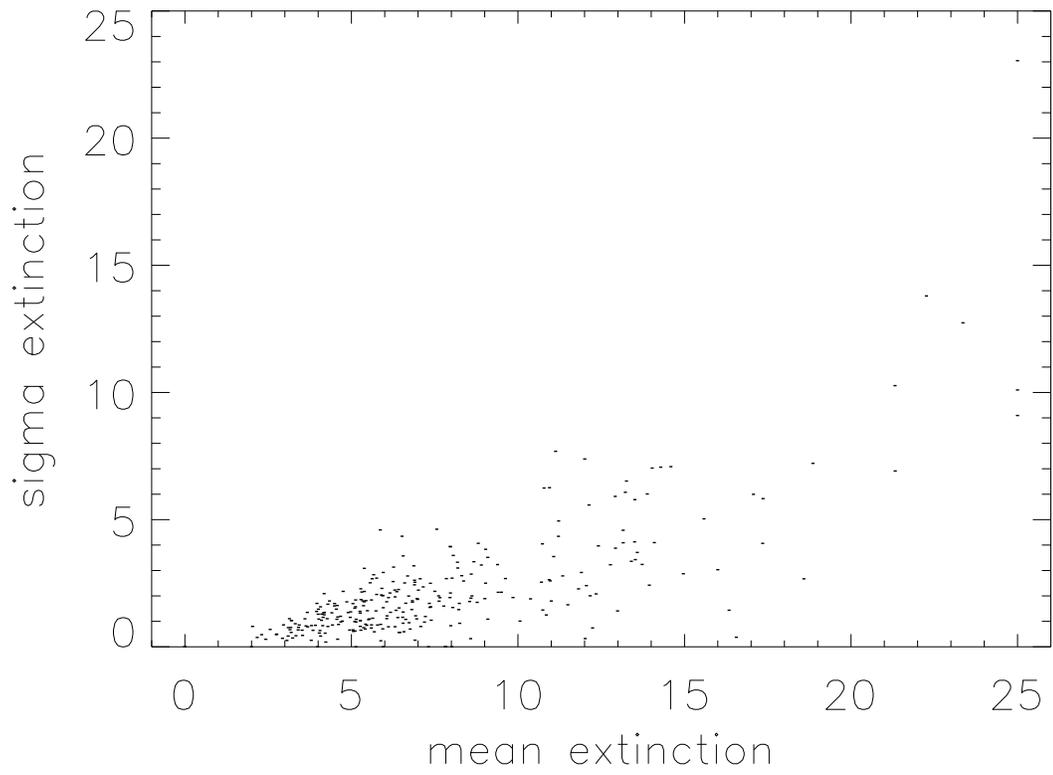}
\caption[]{The same as in fig.2, but from a Log-Normal distribution: now the
observational trend is reproduced.}
\end{figure}

\newpage
\begin{figure}
\centering
\leavevmode
\epsfxsize=1.0
\columnwidth
\epsfbox{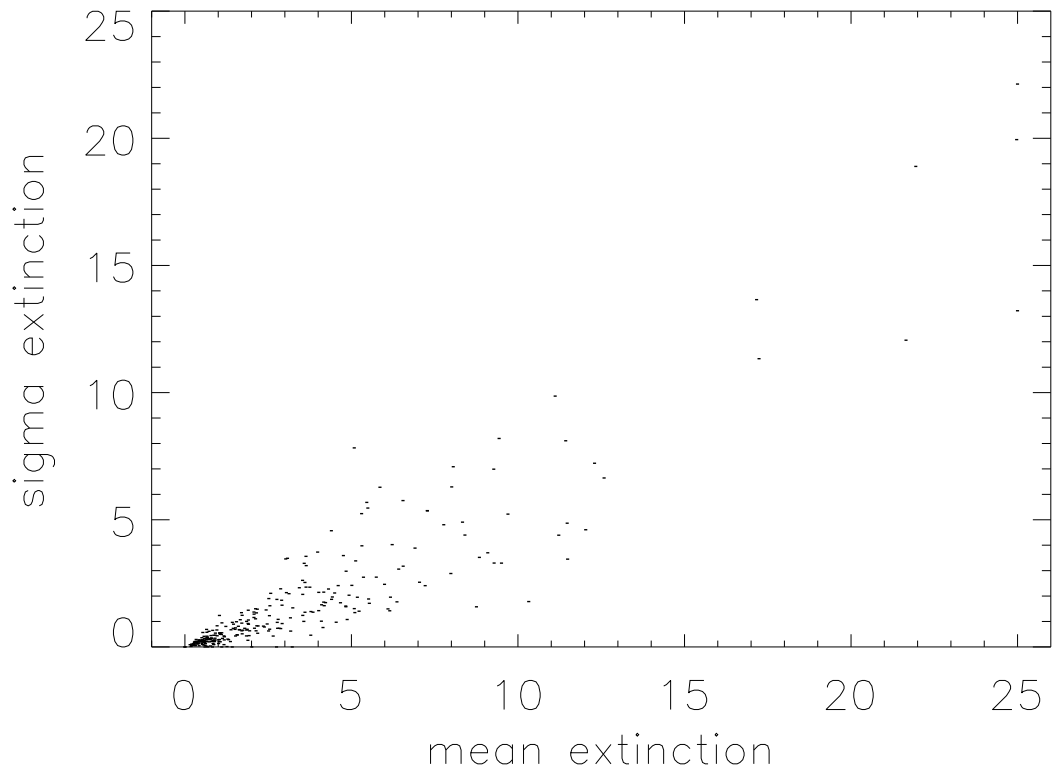}
\caption[]{The same as in fig3, but with larger standard deviation
of the 3-D Log-Normal density distribution.}
\end{figure}

\newpage
\begin{figure}
\centering
\leavevmode
\epsfxsize=1.0
\columnwidth
\epsfbox{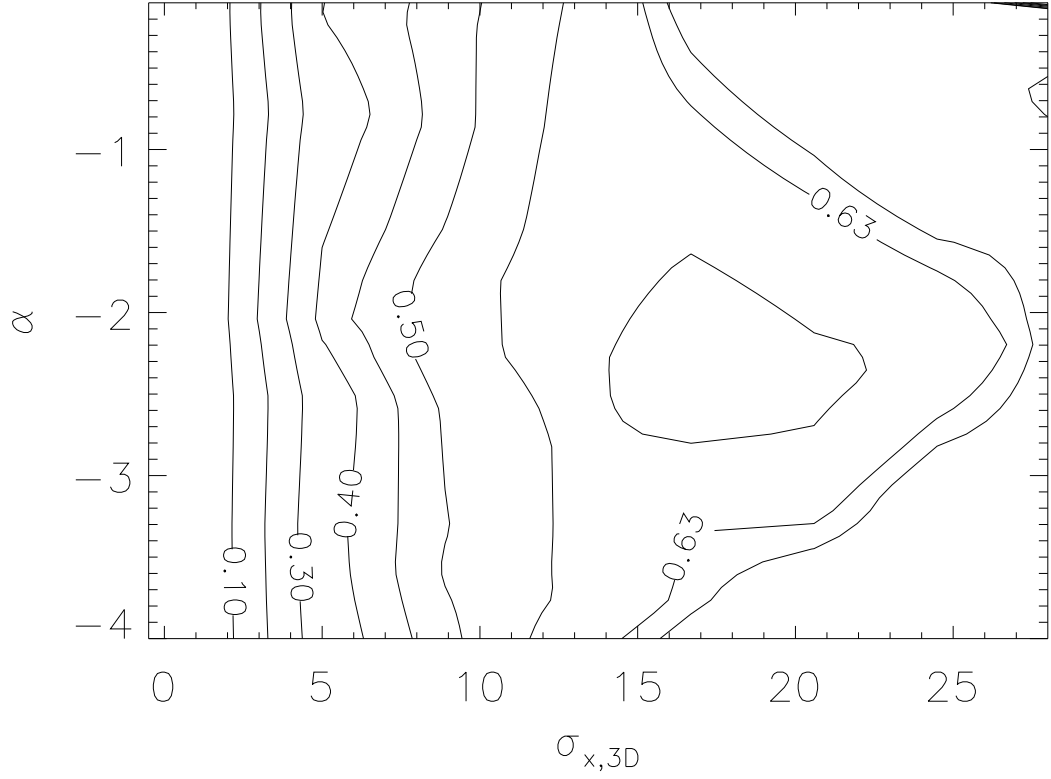}
\caption[]{Contours of constant value of the slope of the
$\sigma-A_{V}$ plot, $C_{r}$. $\alpha$ is the spectral
index of the 3-D density distribution (the power spectrum is
assumed to be a power law), and $\sigma_{x,3D}$ is the
standard deviation of the same distribution. Around the
observed value $C_{r}=0.36$, the plane gives a good constraint
for the 3-D standard deviation, $\sigma_{x,3D}$.}
\end{figure}

\newpage
\begin{figure}
\centering
\leavevmode
\epsfxsize=1.0
\columnwidth
\epsfbox{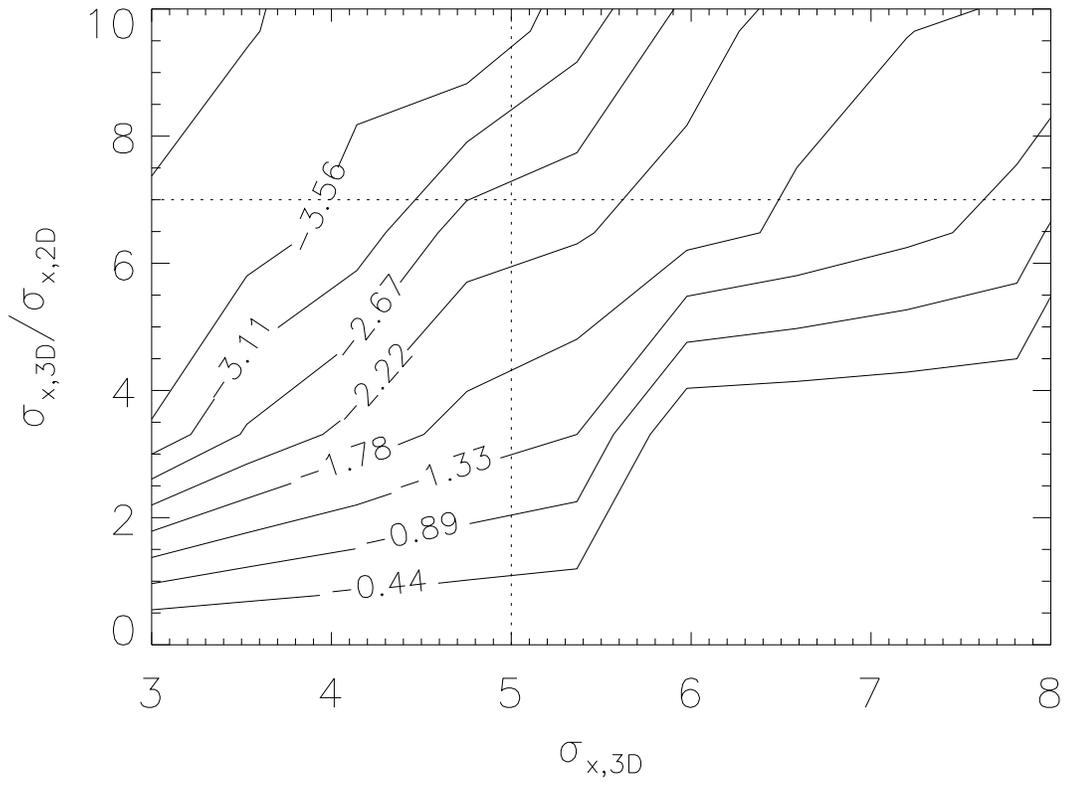}
\caption[]{Contours of constant spectral index. $\sigma_{x,2D}$
and $\sigma_{x,3D}$ are respectively the 2-D and 3-D standard
deviation of the density field.}
\end{figure}

\newpage
\begin{figure}
\centering
\leavevmode
\epsfxsize=1.0
\columnwidth
\epsfbox{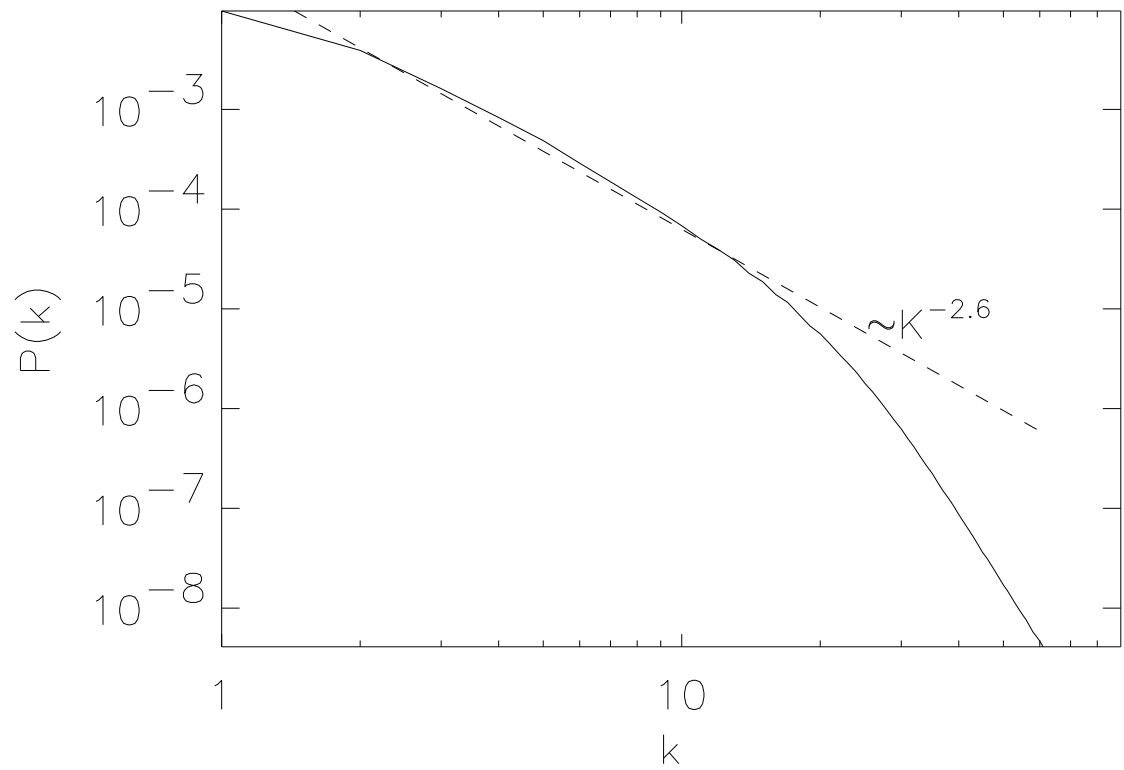}
\caption[]{The power spectrum in 128$^3$ simulations of
supersonic turbulence (Nordlund \& Padoan 1996).
The spectrum is consistent with the index $\alpha=-2.6$
inferred from the observations}
\end{figure}

\end{document}